\documentclass[12pt]{article}
\usepackage{graphicx}

\def\beq{\begin{equation}}
\def\eeq#1{\label{#1}\end{equation}}
\def\eeqn{\end{equation}}

\def\beqa{\begin{eqnarray}}
\def\eeqa#1{\label{#1}\end{eqnarray}}
\def\eeqan{\end{eqnarray}}

\let\bar=\overbar

\def\Dslash{\not{\hbox{\kern-4pt $D$}}}
\def\dslash{\not{\hbox{\kern-2pt $\del$}}}

\def\msb{{\bar{\ssstyle M \kern -1pt S}}}

\def\Title#1{\begin{center} {\Large {\bf #1} } \end{center}}

\begin{document}

\Title{Determination of $c$ and $b$ quark masses}

\bigskip\bigskip


\begin{raggedright}  

{\it Christine Davies\index{Davies, C.}\\
SUPA, School of Physics and Astronomy\\
University of Glasgow\\
Glasgow G12 8QQ, U.K.}
\bigskip\bigskip
\end{raggedright}

\begin{flushleft}
{\it Proceedings of CKM2012, the 7th International Workshop on the CKM 
Unitarity Triangle, University of Cincinnati, USA, 28 September - 2 October 2012}
\end{flushleft}

\section{Introduction}

The determination of quark masses has been transformed in the past few years
by accurate results 
from realistic lattice QCD. This has meant a range of 
new methods for both $b$ and $c$ quarks~\cite{mylat11} which I will describe 
along with results from continuum techniques. 
The recent improvement in Lattice QCD actions 
along with the generation of gluon configurations including 
sea quarks on fine lattices 
has given us
viable methods for handling heavy quarks with relativistic actions. 
This has allowed us for the first time to connect 
the heavy and light sectors through 
accurate determination of ratios of quark masses, such as $m_c/m_s$~\cite{hisqmcms}. 
Such ratios enable us to leverage the accuracy of $c$ and $b$ 
mass determinations into accurate $s$ and light quark masses. 
The comparison of quark masses from different methods 
and formalisms provides a strong test of QCD and the masses themselves 
are needed in calculations for the cross-sections of various 
processes at LHC, for example $H \rightarrow b\overline{b}$. 

\section{Lattice QCD methods}

Lattice QCD calculations have the advantage over the real world of 
having direct access to the parameters of the QCD Lagrangian. These 
are tuned against experimentally well-determined hadron masses, one 
for each quark mass. The masses in the lattice QCD Lagrangian are in 
a lattice regularisation scheme (which depends on the discretisation of 
the QCD Lagrangian used) and at a scale which depends on the inverse lattice 
spacing. To convert them to the standard $\overline{MS}$ scheme at a 
fixed scale must be done either directly, by determining a renormalisation 
constant for the conversion, or indirectly, by calculating on the lattice 
some ultraviolet-finite quantity whose value is known in the continuum 
in terms of the 
$\overline{MS}$ quark mass. 

The direct conversion, $\overline{m}(\mu) = Z(\mu a) m_{latt}(1/a)$,
either calculates $Z$ in lattice QCD perturbation theory or matches 
lattice calculations to a MOM-scheme definition that can then be 
converted to $\overline{MS}$ using continuum QCD perturbation theory. 

A well-developed and accurate indirect method for heavy quarks is 
the current-current correlator method~\cite{allison}. 
Here the continuum limit of 
time-moments of a well-defined lattice QCD heavyonium
correlator is compared to continuum QCD perturbation theory for 
$Q^2$-derivative moments of the heavy quark vacuum polarization 
function. The perturbation theory is known through $\alpha_s^3$ 
for some moments since this method has been developed for 
accurate calculations in the continuum - see, for example~\cite{kuhnupdate}.  
There $m_b$ and $m_c$ are extracted from experimental 
data by isolating 
the $b$ and $c$ quark contributions to 
$\sigma(e^+e^- \rightarrow \mathrm{hadrons})$~\cite{kuhnupdate}. 

Handling $b$ and $c$ quarks in lattice QCD is made 
difficult by
the presence of potentially large discretisation errors 
when the heavy quark mass in lattice units, $m_Qa$, is $\mathcal{O}(1)$. 
These errors can be beaten down by systematically 
improving the discretisation of the Dirac Lagrangian and this is 
the best approach for $c$ quarks. 
For example, the Highly Improved Staggered Quark (HISQ) 
action~\cite{hisq} gives only few \% errors for $c$ physics at 
lattice spacings, $a \approx 0.1 \,\mathrm{fm}$. 
Using this same approach for $b$ physics~\cite{hisqmasses} 
requires very fine lattices and 
extrapolation to the $b$ quark mass from smaller masses where 
discretisation errors are under better control, but has the 
advantage that the same lattice action can be used for all 
5 quarks from $u$ to $b$. Alternatively for $b$ quarks there are a 
number of discretised effective theory approaches 
that avoid large discretisation errors because the $b$ quarks 
are nonrelativistic. The price to be paid, however, is a more 
complicated action that needs more renormalisation.

\begin{figure}[htb]
\begin{center}
\includegraphics[width=0.33\hsize]{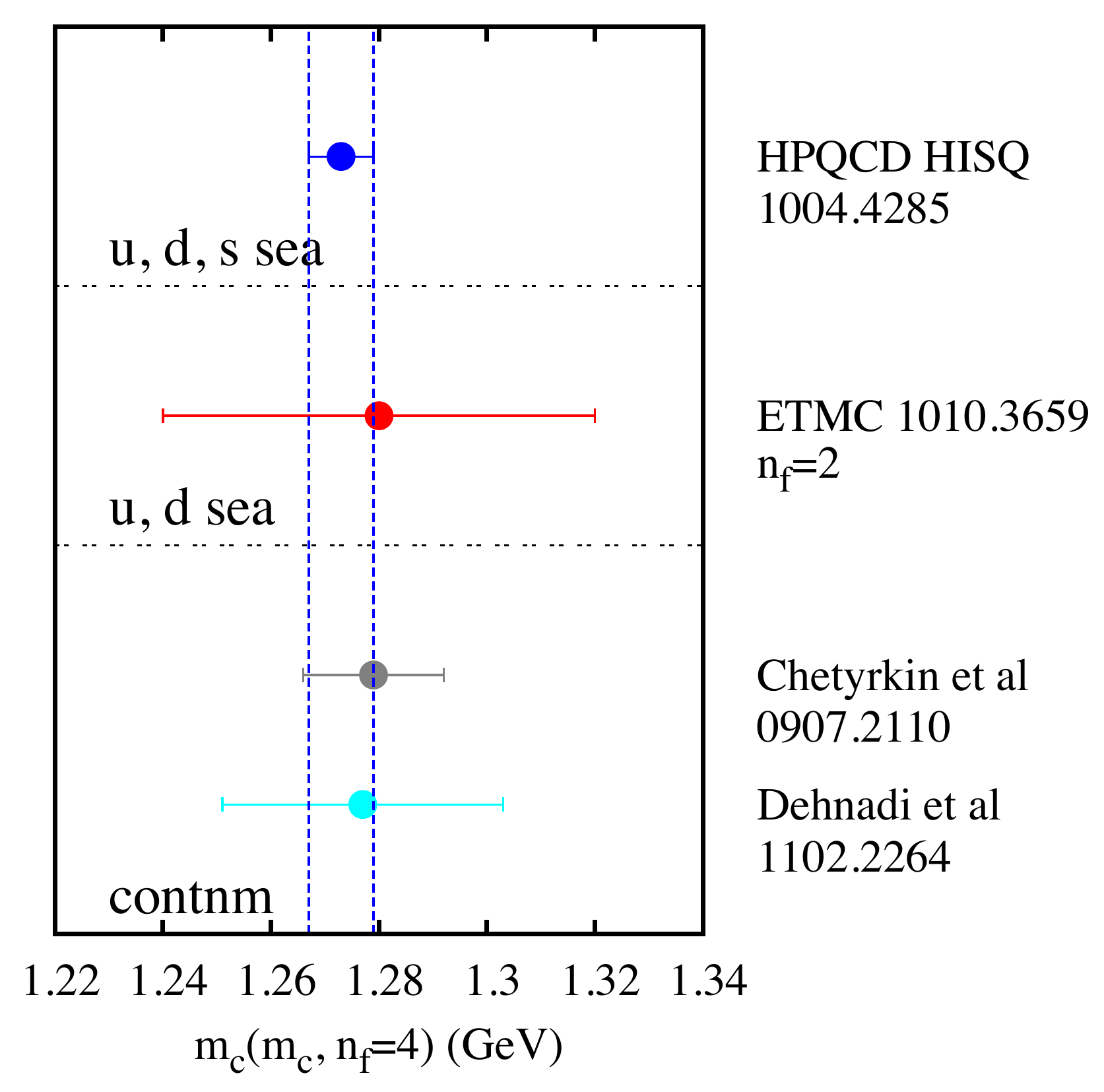}
\includegraphics[width=0.32\hsize]{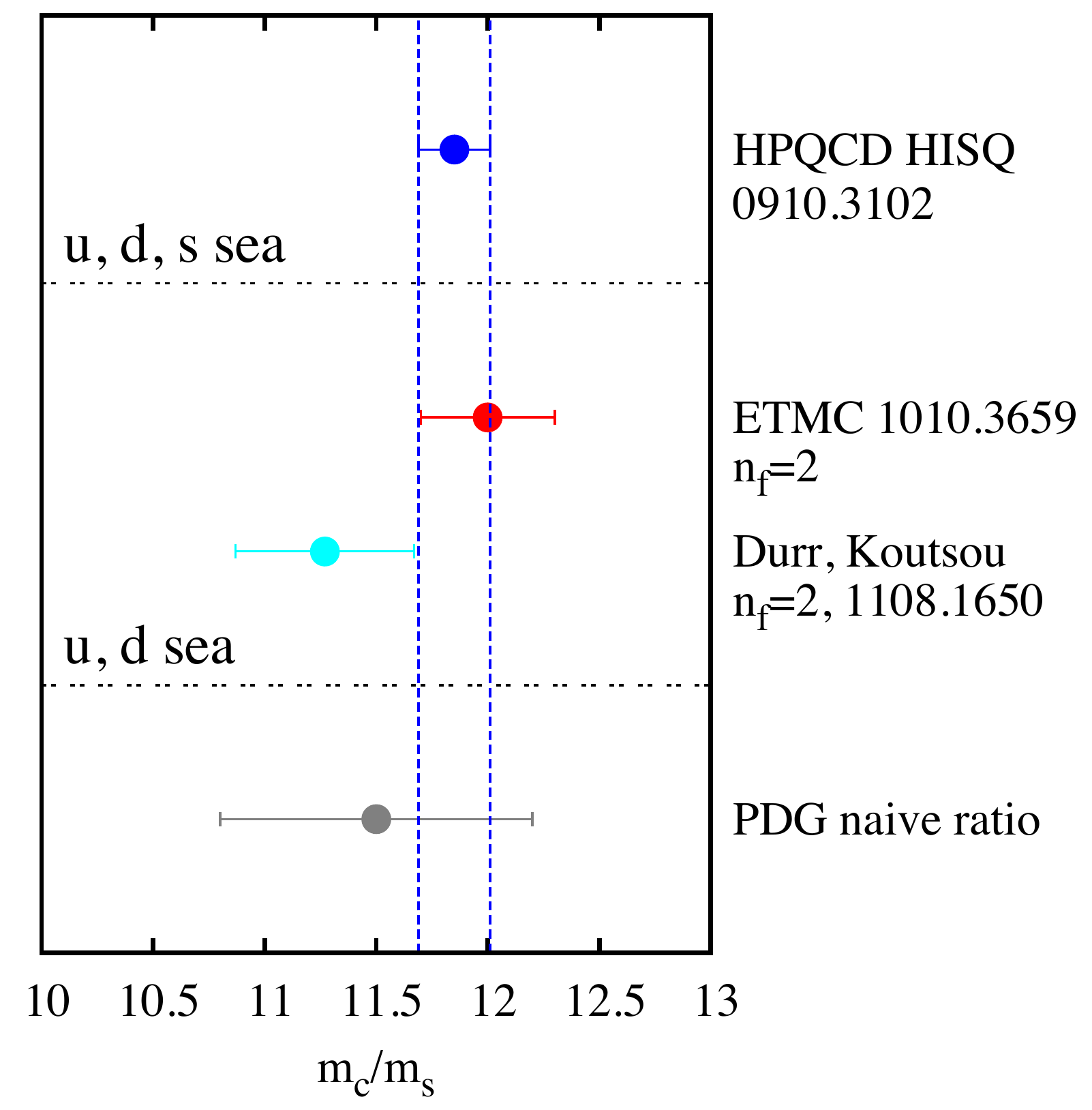}
\includegraphics[width=0.32\hsize]{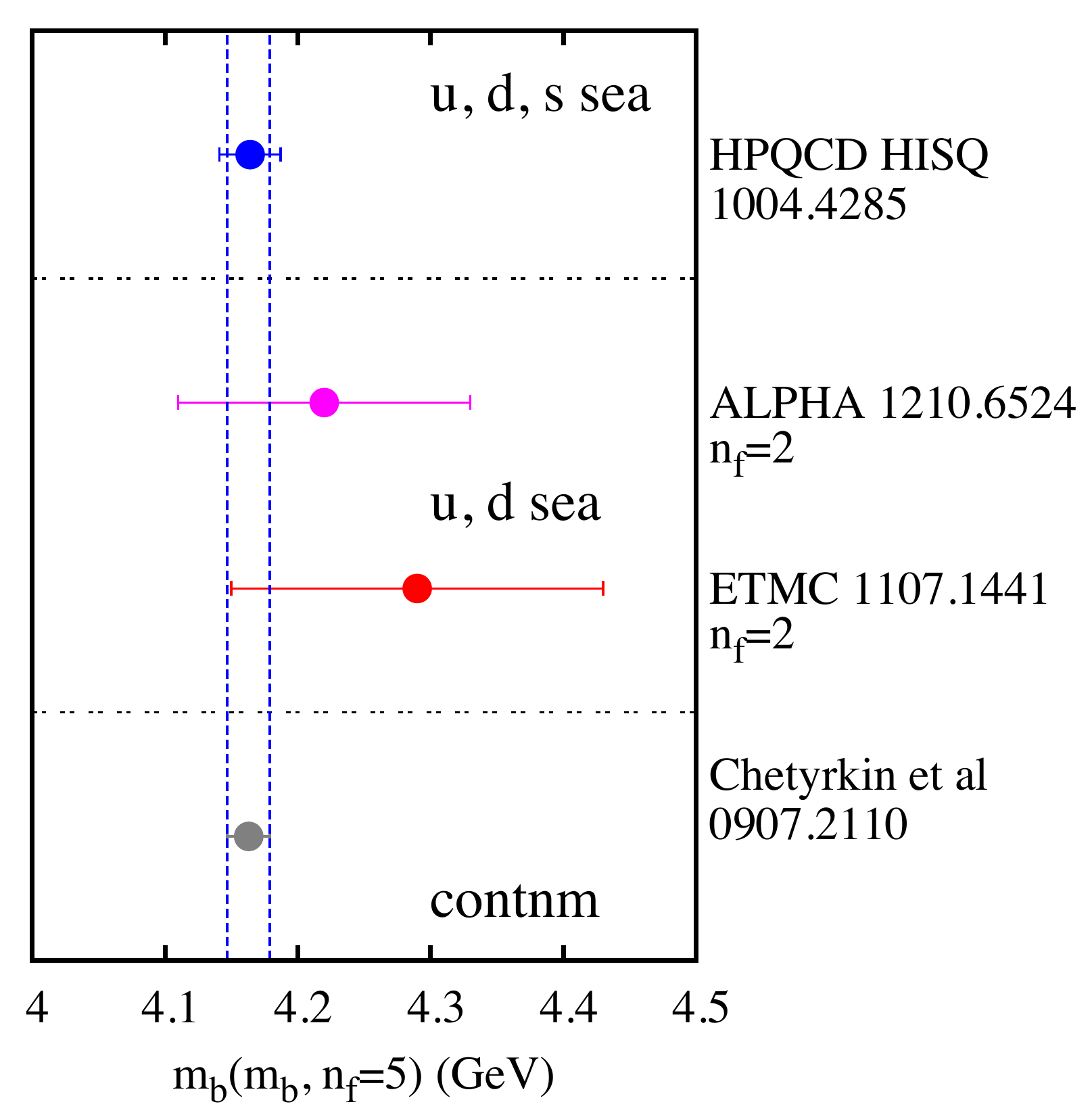}
\caption{Comparison plots for (from left to right): $m_c$, $m_c/m_s$ and $m_b$.
Masses are in the $\overline{MS}$ scheme with a scale equal to the 
mass itself. The ratio is scale-invariant. The lattice QCD results 
marked as `$u,d$ sea' do not include $s$ quarks in the sea and therefore 
may not be comparable to the other numbers. }
\label{fig:comp}
\end{center}
\end{figure}


\section{Results}

Figure~\ref{fig:comp} compares results from lattice QCD and continuum 
determinations for $m_c$ (left) and $m_b$ (right). 
For $m_c$ 1\% errors 
are possible and for $m_b$, 0.5\%. The most accurate values 
are: $\overline{m}_c(\overline{m}_c)=1.273(6) \,\mathrm{GeV}$~\cite{hisqmasses} 
and $\overline{m}_b(\overline{m}_b)=4.163(16) \,\mathrm{GeV}$~\cite{kuhnupdate}.   

\begin{figure}[htb]
\begin{center}
\includegraphics[width=0.48\hsize]{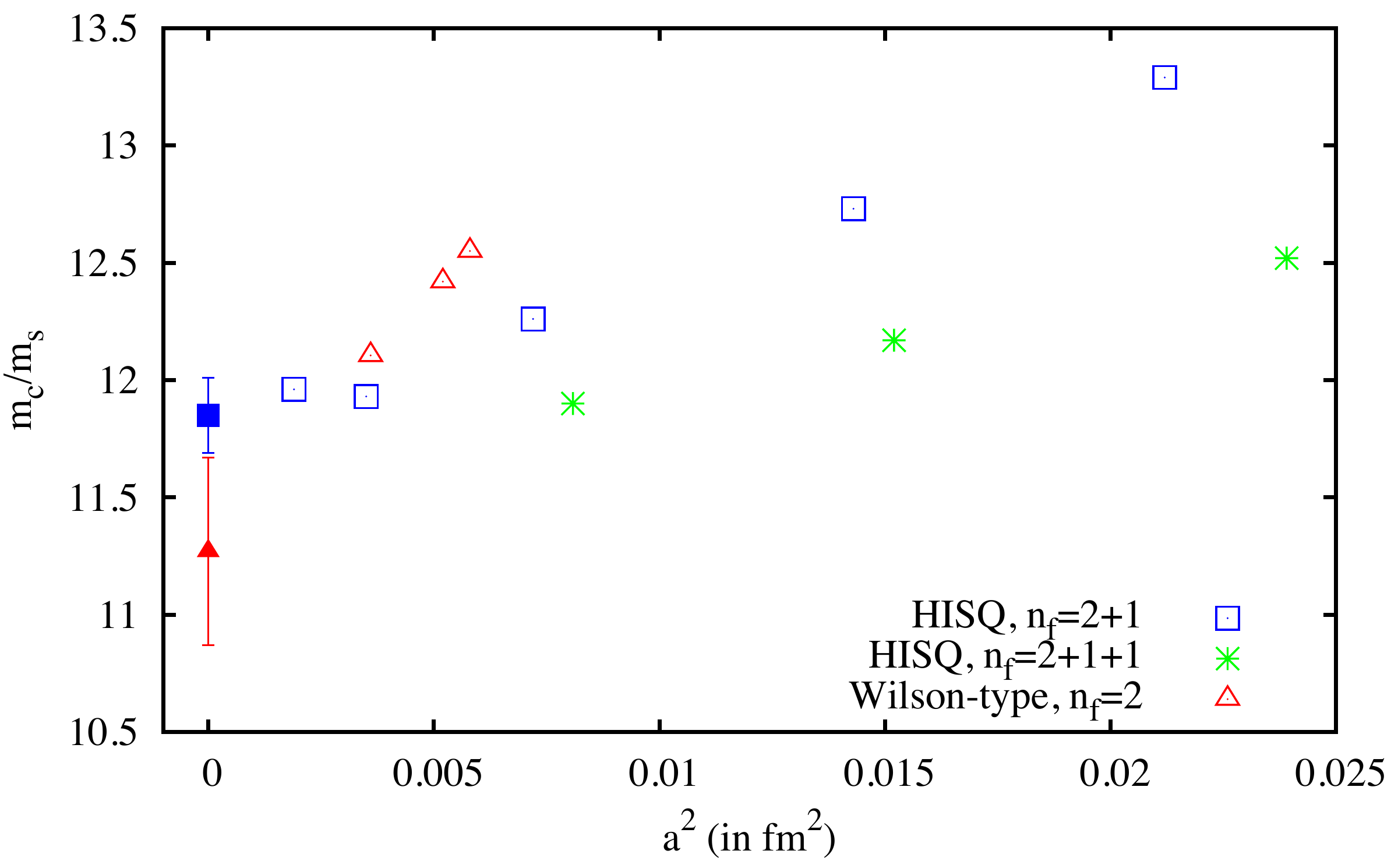}
\includegraphics[width=0.45\hsize]{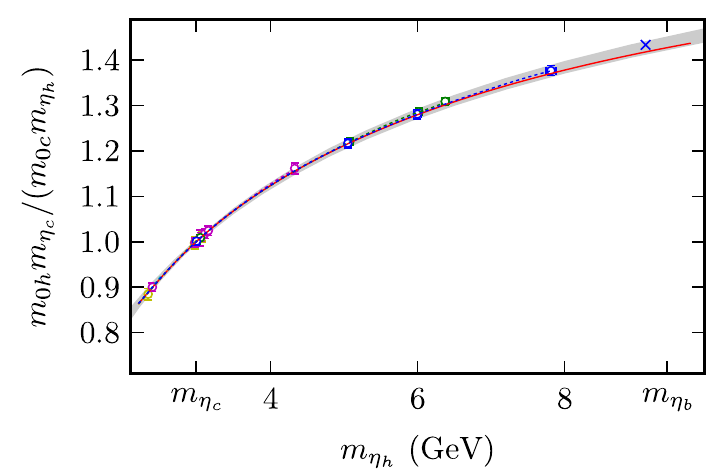}
\caption{ The figure on the left shows the ratio $m_c/m_s$ plotted 
against the square of the lattice spacing for results using 
HISQ quarks on $n_f=2+1$ configurations~\cite{hisqmcms} (blue open squares) and 
Wilson-type quarks on $n_f=2$ configurations~\cite{durrmcms} (red open 
triangles).  
Errors are only shown on the final extrapolated result (filled symbol) 
at $a=0$. 
Green bursts show preliminary HISQ results 
on $n_f=2+1+1$ configurations~\cite{recenthl}.
The figure on the right shows the ratio $m_h/m_c$, divided 
by the ratio of the pseudoscalar meson masses, plotted 
as a function of the pseudoscalar heavyonium mass in the 
range from $m_h=m_c$ to $m_h=m_b$~\cite{hisqmasses}.  
}
\label{fig:rat}
\end{center}
\end{figure}

In the plot for $m_c$, the result from the  
ETM Collaboration~\cite{etmc-mc} is a direct determination using the twisted 
mass quark formalism~\cite{twistedmass} on gluon 
configurations including only $u/d$ quarks in the sea ($n_f=2$) and a $Z$ 
factor determined using the MOM scheme. Note that this result is 
$m_c(n_f=2)$ and so may differ from the other results since there 
is no perturbative way of correcting for missing sea $s$ quarks. 
The other results use the 
current-current correlator method. The HPQCD result~\cite{hisqmasses} uses HISQ quarks 
on gluon configurations including sea $u$, $d$ and $s$ quarks ($n_f=2+1$) 
comparing time-moments of the absolutely normalised 
(in this formalism) pseudoscalar charmonium correlators to continuum 
QCD perturbation theory. The $n_f=3$ quark mass is converted to 
$n_f=4$ perturbatively.  The lower two 
results~\cite{kuhnupdate, hoangmc} are from 
comparing continuum QCD perturbation theory to moments of the $c$ quark 
contribution (effectively via the vector charmonium correlator) 
to $\sigma(e^+e^- \rightarrow \mathrm{hadrons})$. The two results 
differ in their treatment of errors, particularly the perturbative 
error which is handled in both cases by varying the scale at 
which $\alpha_s$ is determined. Both have a significant 
error from the value of $\alpha_s$. The lattice QCD result from 
HPQCD~\cite{hisqmasses} has an advantage here in that $\alpha_s$ is simultaneously 
determined along with $m_c$ (and $m_b$). A particularly accurate 
extraction of $\alpha_s$ is possible from the 
lowest moment of the pseudoscalar correlator, which has no 
explicit mass dependence. In fact the lattice calculation fits 
the 4 lowest moments simultaneously and for a range of masses 
from $c$ to $b$. This allows both a constraint on higher order 
terms in perturbation theory beyond $\alpha_s^3$ and a determination 
of the perturbative error from the effect of including them. 
Moments of the lattice QCD vector correlator can be compared 
directly to the results extracted from 
$\sigma(e^+e^- \rightarrow \mathrm{hadrons})$~\cite{kuhnupdate} and this provides 
a 1.5\% test of QCD~\cite{jpsi} (see also~\cite{etm-cc}). 

The rightmost plot of Figure~\ref{fig:comp} compares results for $m_b$.
The top value from HPQCD/HISQ~\cite{hisqmasses} uses the current-current 
correlator method and is determined in the same analysis as 
$m_c$ above. The bottom result is also from the same continuum 
determination as the $m_c$ above~\cite{kuhnupdate}, 
using $e^+e^-$ data in a different region of $\sqrt{s}$. These 
are the most accurate results. 
The 
two central lattice QCD results use different techniques and so far 
both include only $u/d$ quarks in the sea, so the results 
may not be comparable. The Alpha collaboration~\cite{alpha-mb} uses static 
(infinite mass) quarks but with $1/m_Q$ corrections and determines 
the $b$ mass fom the binding energy of a heavy-light meson using 
nonperturbative step-scaling to determine the energy offset.
The ETM collaboration~\cite{etmc-mb} interpolates the ratio of 
the heavy-light meson mass to the quark mass between relativistic twisted 
mass quarks and static quarks. 
A direct determination of $m_b$ is available from lattice NRQCD on gluon 
configurations that include $u$, $d$, $s$ and $c$ quarks in the sea, but it 
uses a $Z$ factor only 
determined to $\mathcal{O}(\alpha_s)$ in lattice QCD perturbation 
theory so is rather inaccurate~\cite{recentups}.  

Figure~\ref{fig:rat} shows the quark mass ratios $m_c/m_s$ and 
$m_b/m_c$ determined from lattice QCD calculations. The ratio 
of lattice QCD masses 
translates directly to the ratio of 
$\overline{MS}$ masses (at the {\it same} scale) because 
the $Z$ factors cancel when the same 
formalism is used for both quarks. This is then a fully nonperturbative 
determination and can be done with an accuracy of 1\%. 

The leftmost plot
shows results for $m_c/m_s$ as a function of $a^2$ for HISQ~\cite{hisqmcms} and 
Wilson-type quarks~\cite{durrmcms}. The improved $a$-dependence of HISQ quarks 
is evident, particularly for new results 
on $n_f=2+1+1$ configurations that include a further improved 
gluon action and HISQ quarks in the sea~\cite{recenthl}. 
A comparison of different lattice QCD results for $m_c/m_s$ is 
given in the middle plot of Figure~\ref{fig:comp}, compared to the 
ratio obtained from the Particle Data Tables~\cite{pdg} running $m_c$ and $m_s$ 
to the same scale. The best, and only $n_f=2+1$, result 
is $m_c/m_s =11.85(16)$  
from HPQCD~\cite{hisqmcms}. It can be used with an accurate value for $m_c$ to 
obtain a 1\% accurate $m_s$~\cite{hisqmasses}. 

The rightmost plot of Figure~\ref{fig:rat} shows HISQ 
results for $m_h/m_c$ over the 
range of heavy quark masses from $c$ to $b$. Very little $a$-dependence 
is seen and a 1\% accurate result at the $b$ can be 
obtained: $m_b/m_c = 4.49(4)$~\cite{hisqmasses}. This agrees well with 
the ratio of masses obtained from the current-current correlator method 
and provides a good nonperturbative test of that method.

\end{document}